\documentclass[aps, prl, twocolumn, superscriptaddress, amsmath, 
]{revtex4-2}

\usepackage{graphicx}
\usepackage{amsfonts}
\usepackage{amssymb}
\usepackage{amsmath}
\usepackage[normalem]{ulem}
\usepackage{hyperref}
\hypersetup{
    colorlinks=true,       
    linkcolor=blue,          
    citecolor=blue,        
    filecolor=blue,      
    urlcolor=blue           
}
\usepackage{makecell}
\usepackage{tabularx}
\usepackage{array}

\newcommand{\Bo}[1]{\noindent{\bf #1}}
\newcommand{\ket}[1]{\vert#1\rangle}

\newcommand{\be}{\begin{equation}}
\newcommand{\ee}{\end{equation}}
\newcommand{\ba}{\begin{array}}
\newcommand{\ea}{\end{array}}
\newcommand{\bea}{\begin{eqnarray}}
\newcommand{\eea}{\end{eqnarray}}

\def\6{{\langle}}
\def\9{{\rangle}}
\def\half{{\tfrac{1}{2}}}
\renewcommand{\figurename}{Fig.}

\usepackage{xcolor}

\begin{document}
	
\title{Controlling wave--particle duality with quantum entanglement}

\author{Kai Wang}
\affiliation{National Laboratory of Solid-state Microstructures, School of Physics, Collaborative Innovation Center of Advanced Microstructures, Nanjing University, Nanjing 210093, China}
	
\author{Daniel R. Terno}
\affiliation{Department of Physics and Astronomy, Macquarie University, Sydney, New South Wales, Australia}
\affiliation{Institute for Quantum Science and Engineering, Department of Physics, Southern University of Science and Technology, Shenzhen 518055, China }
	
\author{\v{C}aslav Brukner}
\affiliation{Vienna Center for Quantum Science and Technology (VCQ), University of Vienna, Faculty of Physics, Vienna, Austria}
\affiliation{Institute for Quantum Optics and Quantum Information (IQOQI), Austrian Academy of Sciences, Vienna, Austria}
	
\author{Shining Zhu}
\affiliation{National Laboratory of Solid-state Microstructures, School of Physics, Collaborative Innovation Center of Advanced Microstructures, Nanjing University, Nanjing 210093, China}
	
\author{Xiao-Song Ma}
\email{Xiaosong.Ma@nju.edu.cn}
\affiliation{National Laboratory of Solid-state Microstructures, School of Physics, Collaborative Innovation Center of Advanced Microstructures, Nanjing University, Nanjing 210093, China}
	
\date{\today}
	
\begin{abstract}
Wave--particle duality and entanglement are two fundamental characteristics of quantum mechanics. All previous works on experimental investigations in wave--particle properties of single photons (or single particles in general) show that a well-defined interferometer setting determines a well-defined property of single photons. Here we take a conceptual step forward and control the wave-particle property of single photons with quantum entanglement. By doing so, we experimentally test the complementarity principle in a scenario, in which the setting of the interferometer is not defined at any instance of the experiment, not even in principle. To achieve this goal, we send the photon of interest (S) into a quantum Mach--Zehnder interferometer (MZI), in which the output beam splitter of the MZI is controlled by the quantum state of the second photon (C), who is entangled with a third photon (A). Therefore, the individual quantum state of photon C is undefined, which implements the undefined settings of the MZI for photon S. This is realized by using three cascaded phase stable interferometers for three photons. There is typically no well-defined setting of the MZI, and thus the very formulation of the wave--particle properties becomes internally inconsistent.
\end{abstract}

\maketitle
Wave--particle duality is a fundamental property of quantum systems. While its historic origins reach back at least to the seventeenth century~\cite{wheaton2009wave},
 precise content has been formulated in the early years of quantum mechanics~\cite{wheeler1984quantum}: a single quantum system can be described accurately only by combining classically incompatible concepts of particles and waves\cite{grangier1986experimental}. This deceivingly simple formulation lies at the heart of one of the oldest debates on quantum foundations.

In the 1970s, Wheeler proposed the seminal delayed-choice gedanken experiment to rule out certain naive interpretations of complementarity~\cite{WHEELER19789,wheeler1984quantum}. His proposal is illustrated in Fig.~1(\textit{A}). A photon is sent through a two-path Mach--Zehnder interferometer (MZI), where one of the paths is equipped with a tunable phase shifter, $\varphi$. The two paths of MZI are recombined (or not) at a second beam splitter (BS) before detection. In either case, we randomly detect a click in only one of the two detectors terminating the arms. If the second BS is present, then for a generic phase we reconstruct the phase-dependent interference fringes by collecting enough detection events, indicating that the photon behaved as a wave, traveling through both arms of the MZI. If the second BS is absent, the statistics depends only on the splitting ratio of the first BS, revealing particle-like behaviour of the photon. In Wheeler's delayed-choice experiment, one chooses whether or not to insert the second BS when the photon is already inside the interferometer. The particle or the wave behaviour of the single photon therefore could not have been determined at the first BS of the MZI.

This gedanken experiment and its realizations~\cite{alley1986results, PhysRevA.35.2532, Baldzuhn1989, jacques2007experimental,PhysRevA.85.032121,Manning2015,PhysRevLett.120.190401,PhysRevA.100.022111,PhysRevA.100.012114,PhysRevA.100.012115}
show that the behaviour of a single photon (or a quantum system in general) in a single experiment depends only on the entire experimental context, not on the temporal order of individual events (for recent reviews, see refs~\cite{Shadbolt2014,RevModPhys.88.015005}).

\begin{figure}
\includegraphics[width=8.5cm]{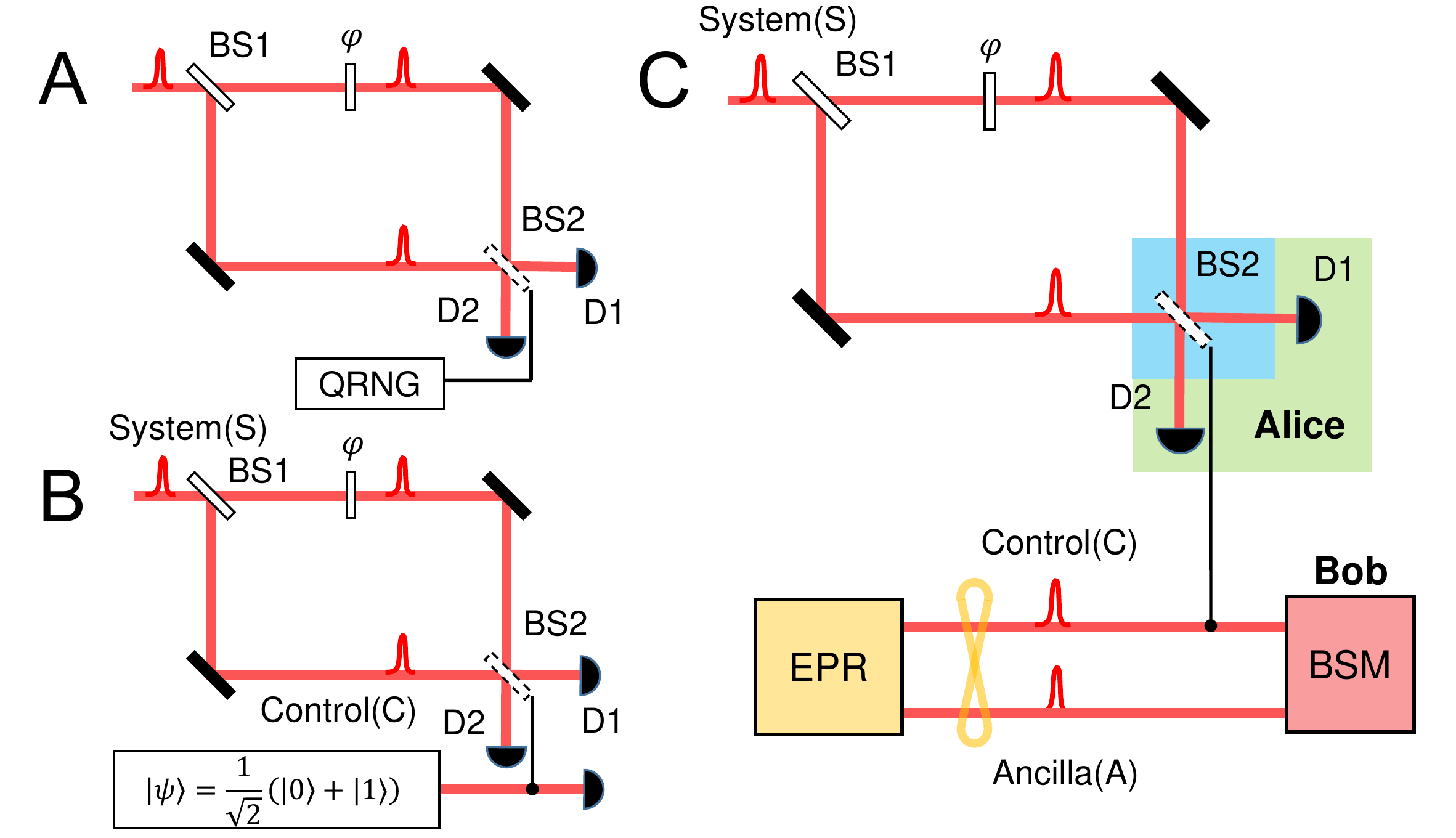}
\caption{The advances of delayed-choice experiments. (\textit{A}). Wheeler's delayed-choice experiment. To insert or remove the output beam splitter (BS2) is decided by a quantum random number generator (QRNG) and delayed until after the single photon passed through BS1. One measures the interference fringe or the second-order correlation function, conditional on the classical control bit value, to quantify the wave-particle properties of single photons. (\textit{B}). Quantum delayed-choice experiment. BS2 is implemented with a quantum beam splitter, whose setting is determined by the superposition state of a control photon (C). One measures the correlation functions between the system and ancilla photons, and violates Bell inequality to quantify the wave-particle properties of single photons. (\textit{C}). Entanglement--controlled wave--particle duality experiment. The system photon (S) enters into an interferometer composed by BS1, a phase shifter ($\varphi$) and a quantum-controlled Hadamard gate (BS2, blue). The settings of CH are determined by the entanglement of photons C and A, which are generated from an Einstein--Podolsky--Rosen (EPR) entangled-photon-pair source. After the CH operation, the quantum entanglement of photons C and A controls the superposition of the wave and particle states of photon S. Alice performs single-photon measurement (green) on photon S, whereas Bob performs a Bell-state measurement (BSM, red) on photons C and A. } 
\end{figure}


Later, Ionicioiu and Terno proposed the quantum delayed-choice experiment (QDC). In QDC, the classical random bit used to control the state of MZI is replaced by a qubit (control photon C), as shown in Fig.1(B). When the qubit is in the superposition state of 0 and 1, the MZI would be in the quantum superposition state of open and closed. Note that the first experiment to show the wave-particle duality of one photon being controlled by a second photon has been demonstrated in~\cite{PhysRevLett.92.190402}. Based on this configuration, framing the original discussion in terms of hidden variables (HV) \cite{PhysRevLett.107.230406} not only inspired the design of novel experiments \cite{ionicioiu2014wave}, but revealed that any theory that assumes weak determinism, $\lambda$-independence (independence of hidden variables on the conducted experiments), and objectivity is inconsistent \cite{PhysRevLett.114.060405}. Precise definitions of different types of determinism and independence that are the building blocks of the paradoxes of quantum mechanics can be found in \cite{brandenburger2008a}. We provide a brief summary of them in \textit{Appendix}, Sec.~I. By objectivity, we refer to the assumption that two types of statistics that are demonstrated by some system in two different experimental set-ups are a manifestation of some intrinsic properties.

These theoretical studies triggered a recent resurgence of interest in wave--particle duality \cite{RevModPhys.88.015005}. In particular, it was shown that wave--particle duality of a photon can be quantum-coherently controlled by entangling it with another photon~\cite{PhysRevLett.107.230406,ionicioiu2014wave}. This concept has been demonstrated in various systems~\cite{tang2012realization, peruzzo2012quantum, kaiser2012entanglement, PhysRevLett.115.260403, liu2017twofold, PhysRevA.85.022109, PhysRevA.92.022126}. Moreover, a recent experiment showed that the wave--particle entanglement can be established between two photons~\cite{rab2017entanglement}.

\begin{figure}
\includegraphics[width=9cm]{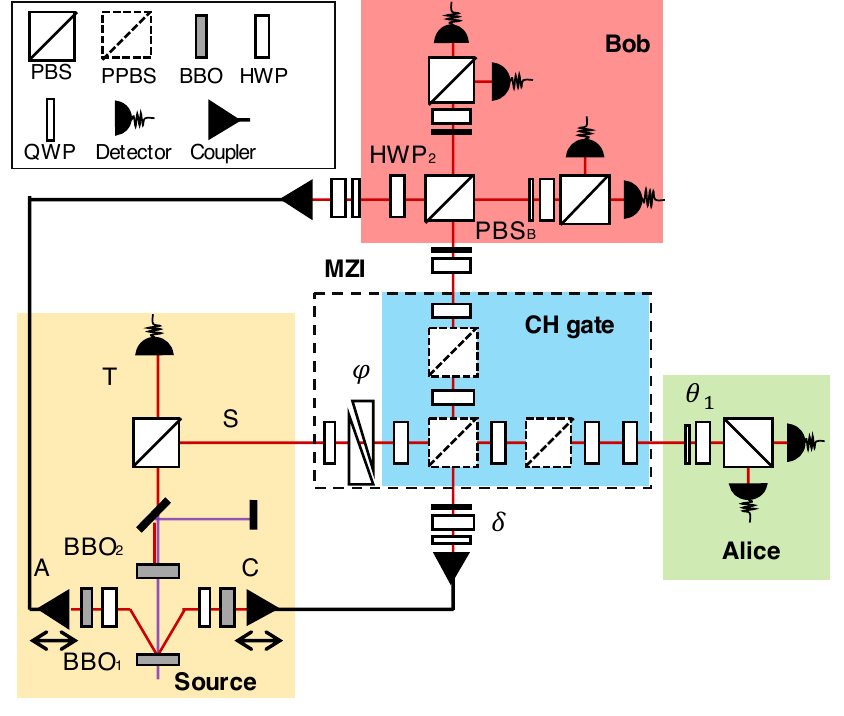}
\caption{Experimental setup.  Entangled photons C and A are generated in the $\ket{\psi^{+}}$ state from a $\beta$-barium borate crystal (BBO1). From BBO2, we detect photon T and herald the existence of photon S. We then send photon S through the polarization Mach-Zehnder interferometer (MZI), which consists of one half-wave plate (HWP), a Soleil--Babinet compensator (SBC) and a quantum-CH gate. Alice performs polarization measurements on photon S with two wave plates and a polarizing beam splitter (PBS) which project the polarization of photon S along $\theta_{1}$. Photons C and A are sent to Bob, who employs a Bell-state analyzer to project these two photons onto a coherent superposition of $\ket{\phi^{-}}$ and $\ket{\psi^{+}}$. HWP$_{2}$ tunes the superposition of the bipartite states of photons C and A.}
\end{figure}

These previous demonstrations typically focus on measuring the statistics of the photon of interest (system photon) conditionally on the measurement of a second photon (control photon) \cite{kaiser2012entanglement, peruzzo2012quantum, Ma1221} or a second degree of freedom of the same system photon~\cite{tang2012realization}. One exception is the recent realization of a non-local quantum delayed-choice experiment with three photons (the system, the control and an ancilla photon) ~\cite{wang2019quantum}. A Hadamard gate realizes the second BS, controlled by the ancilla photon. By performing measurements set by the active and independent choices on the ancilla photon at a space-like separated region, the Hadamard gate is prepared via the control photon and enables the demonstration of wave--particle duality of the system photon. In ref.~\cite{wang2019quantum}, the setting of the Controlled Hadamard (CH) gate is determined by the `classical' correlation between photons C and A, which is defined by the individual measurements of photons C and A. By doing so, one excludes the possibility that the choice of the measurement can communicate with the system photon, which then can adapt its behaviour according to the choice and reproduce what quantum physics predicts.

Complementarity principle states that the mutually exclusive settings of the experiment apparatus determines the complementary properties of the quantum particle we reveal in any specific run of the experiment. All previous works, realized by using either a classical state, or a well-defined single-qubit quantum state via state preparation/measurement, showed that given various but well-defined settings of the interferometer, including open, close, the  classical mixture of open-or-close, and even coherent superposition of the open-and-close status, complementarity principle is valid. Therefore, in any run of an experiment, a property of the photon is well-defined after the measurements, being a wave, or a particle, a classical mixture of wave and particle, or a quantum superposition of wave and particle.

In this work, we take a conceptual step forward and experimentally test the complementarity principle in a scenario, in which the setting of the interferometer is not defined at any instance of the experiment, not even in principle. To make this possible, we send the photon of interest, photon S, into the quantum MZI, in which the output beam splitter of the MZI is controlled by the quantum state of photon C. Photon C is one part of a maximally entangled bipartite state (formed by photons C and A), as shown in Fig.1(\textit{C}). Therefore, the individual quantum state of photon C is undefined, which implements the undefined settings of the MZI for photon S. Even after the measurements of photons, the settings of the MZI is not defined.  When experimentally obtained correlations violate the Bell's inequality between the entangled states of photons C and A and the state of photon S, we confirm the quantum correlations between these two parts. Thereby, we control the wave--particle duality of S via entangled state of C and A. Please see \textit{Appendix} for the summary of recent experiments on delayed-choice experiments. Ref. \cite{PhysRevLett.120.190401} strongly restricts plausibility of retrocausal interpretations of the delayed choice experiments. In this work, we do not assume any form of retrocausality.

\section*{Theory and experimental setup}
\begin{figure*}{}
\includegraphics[width=17.8cm]{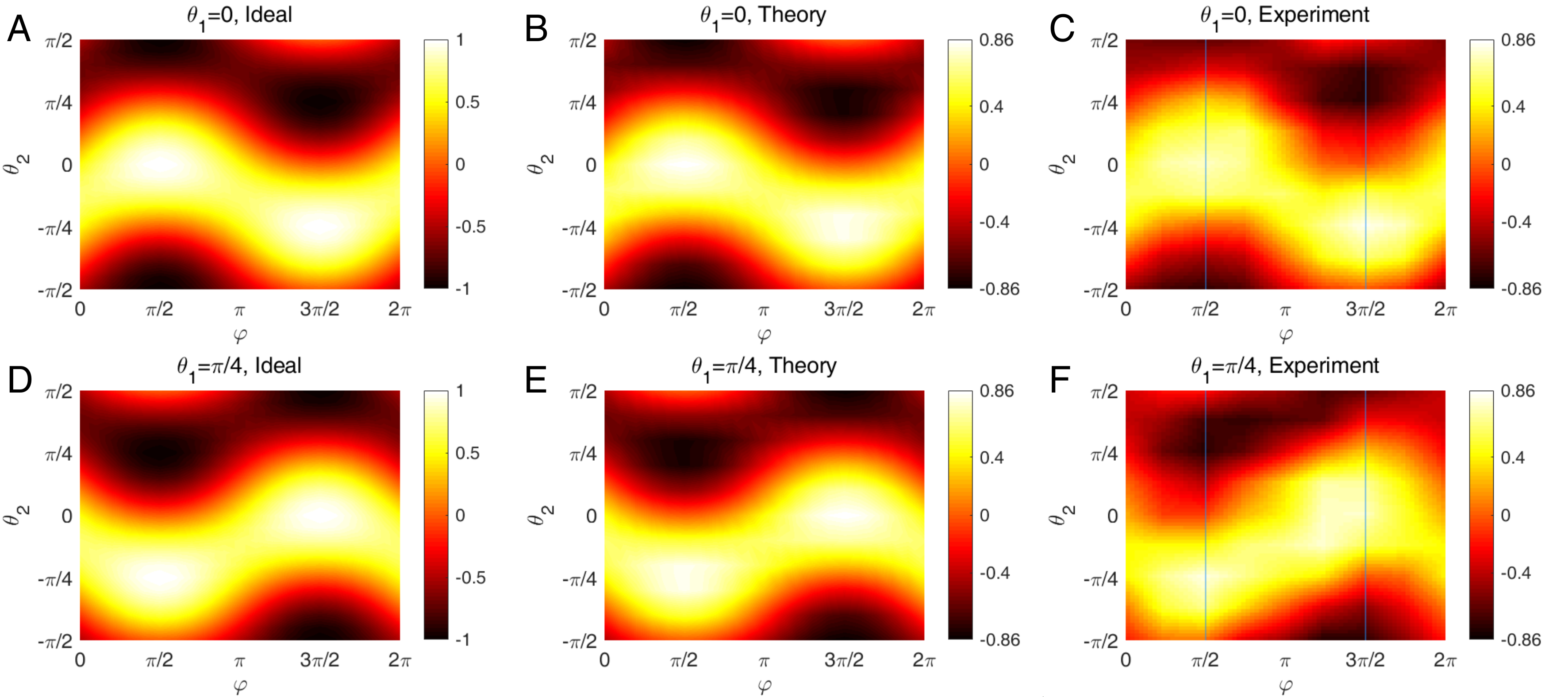}
\caption{ Correlation functions $E(\theta_{1},\theta_{2},\varphi)=\frac{N^{++}-N^{+-}-N^{-+}+N^{--}}{N^{++}+N^{+-}+N^{-+}+N^{--}}$ of final state $\ket{\Psi^{f}(\varphi)}$.
(\textit{A})/(\textit{B}) Ideal and theoretical results of Bob's measurement base $\theta_{1}=0$. $E(\theta_{2},\varphi)=\frac{1}{2}[\cos{2\theta_{2}}-\sin{2\theta_{2}}+\sin{\varphi}(\cos{2\theta_{2}}+\sin{2\theta_{2}})]$.
 (\textit{D})/(\textit{E}) Ideal and theoretical results of Bob's measurement base $\theta_{1}=\frac{\pi}{4}$.
  $E(\theta_{2},\varphi)=\frac{1}{2}[\cos{2\theta_{2}}-\sin{2\theta_{2}}-\sin{\varphi}(\cos{2\theta_{2}}+\sin{2\theta_{2}})]$. In (\textit{C})/(\textit{F}), we experimentally perform a two-dimensional scan with nine values of $\varphi$ and nine values of $\theta_{2}$,
  equally distributed from 0 to $2\pi$ and $-\pi/2$ to $\pi/2$, respectively. The vertical blue lines indicate the measurement settings shown in Fig.3. The intermediate values between each step are linearly interpolated and plotted accordingly. Error bars in (\textit{C}) and (\textit{F}) are derived from Poissonian
  statistics and range from 0.0137 to 0.0344 and 0.0143 to 0.0309, respectively.}
\end{figure*}

The conceptual scheme of our experiment is shown in Fig.~1(\textit{C}). The initial state of the three photons S, C and A is
\begin{equation}
\ket{\psi^{i}_{SCA}}=\ket{V}_{S}\otimes\frac{\ket{HV}_{CA}+e^{i\delta}\ket{VH}_{CA}}{\sqrt{2}},
\end{equation}
where $\ket{H}$, $\ket{V}$ are horizontal and vertical polarization states, respectively. Photon S is vertically polarized and is sent through the MZI made from a Hadamard gate (H), a phase shifter ($\varphi$) and a quantum-controlled Hadamard gate (CH). An entangled photon pair consisting of photons C and A are prepared from an Einstein--Podolsky--Rosen (EPR) source, with the relative phase $\delta$ between the $\ket{HV}_{CA}$ and $\ket{VH}_{CA}$ terms. After local polarization rotations and fixing $\delta$ at $\frac{\pi}{4}$, we have the final state of these three photons as shown in Eq. (2)

\begin{figure*}[bt!]
\begin{align}
\nonumber \ket{\psi^{f}}=&\frac{1}{2}[\ket{p}_S(\ket{\phi^{-}}_{CA}-i\ket{\psi^{+}}_{CA})+e^{i\pi/4}\ket{w}_{S}(\ket{\phi^{-}}_{CA}+i\ket{\psi^{+}}_{CA})]\\
\nonumber =&\frac{1}{\sqrt{2}}[\frac{e^{i\frac{\pi}{4}}\ket{w}_{S}+\ket{p}_{S}}{\sqrt{2}}\ket{\phi^{-}}_{CA}+i\frac{e^{i\frac{\pi}{4}}\ket{w}_{S}-\ket{p}_{S}}{\sqrt{2}}\ket{\psi^{+}}_{CA}]\\
 =&\frac{1}{2}[(e^{i(\theta+\frac{\pi}{4})}\ket{w}+e^{-i\theta}\ket{p})_{S}(\cos{\theta}\ket{\phi^{-}}+\sin{\theta}\ket{\psi^{+}})_{CA}+(e^{i(\theta-\frac{\pi}{4})}\ket{w}+i e^{-i\theta}\ket{p})_{S}(\sin{\theta}\ket{\phi^{-}}-\cos{\theta}\ket{\psi^{+}})_{CA}]
\end{align}
\end{figure*}

where statistics of the states $\ket{p}=\frac{\ket{H}-e^{i\varphi}\ket{V}}{\sqrt{2}}$ and $\ket{w}=e^{i\varphi/2}(-i\sin{\frac{\varphi}{2}}\ket{H}+\cos{\frac{\varphi}{2}}\ket{V})$ show the amplitude independence and dependence on the phase shift, respectively. In the polarization representation, the four Bell states are $\ket{\psi_{CA}^{\pm}}=(\ket{HV}_{CA}\pm\ket{VH}_{CA})/\sqrt{2}$ and $\ket{\phi_{CA}^{\pm}}=(\ket{HH}_{CA}\pm\ket{VV}_{CA})/\sqrt{2}$ for photons C and A.

According to Eq. (2), we control the wave-particle properties of photon S with entangled states of photons C and A as a total system. By doing so, we can probe the particle behaviour of photon S by projecting photons C and A onto a superposition of two Bell states, $(\ket{\phi^{-}}_{CA}-i\ket{\psi^{+}}_{CA})/\sqrt{2}$. Similarly, we can choose to probe the wave behaviour of photon S by projecting photons C and A onto a superposition of two Bell states with phase $\pi/2$, $(\ket{\phi^{-}}_{CA}+i\ket{\psi^{+}}_{CA})/\sqrt{2}$. We can also create a superposition of the wave and particle states of photon S by projecting photons onto $\ket{\phi^{-}}_{CA}$ or $\ket{\psi^{+}}_{CA}$, enabled by the entanglement between individual states of photon S and bipartite states of photons C and A. The outcome of this Bell-state measurement (BSM) corresponds to one of the four Bell states ($\ket{\psi_{CA}^{\pm}}$ and $\ket{\phi_{CA}^{\pm}}$). In our setting, only two of these four outcomes can be obtained: $\ket{\phi_{CA}^{-}}$ and $\ket{\psi_{CA}^{+}}$. With this scheme, we violate a Clauser--Horne--Shimony--Holt (CHSH)--Bell inequality \cite{PhysRevLett.23.880} with the individual quantum state of photon S and the joint state of photons C and A. Although the contexts are different, this concept is related to entangled entanglement~\cite{PhysRevA.54.1793, PhysRevLett.97.020501}. In this work, we focus on controlling the superposition between the particle and wave state of photon S with the superposition of Bell states of photons C and A, as shown by the third line of Eq. (2). Note that one can also investigate the pure particle or wave nature of photon S by projecting photon C and A into $\ket{\phi^-}-i\ket{\psi^+}$ or $\ket{\phi^-}+i\ket{\psi^+}$ , which has not been shown here due to the particular design of our BSM device. However, this does not impact our conclusion, because these controls are equivalent which can be seen in line 1 and 3 of Eq. (2).

The current setup enables the measurements by Bob in the state of $\Phi_{CA}(\theta_{2})=\cos\theta_{2} \ket{\phi^{-}}_{CA}+\sin\theta_{2}\ket{\psi^{+}}_{CA}$,
which leads to Alice to have a wave or a particle state or its superposition, depending on the outcome of Bob. A tunable BSM changes the amplitudes of $\ket{\phi^{-}}_{CA}$ and $\ket{\psi^{+}}_{CA}$ by changing $\theta_{2}$. In order to achieve this, we use of three cascaded interferometers for three photons, which is among the most complex experiments in the context of delayed-choice experiment to the best of our knowledge. This complexity of the experimental setup allows us to implement a quantum-controlled interferometer with one first-order interference for S and one second-order interference for SC. The third interferometer is the second-order interference based on HOM effect, realizing the required BSM for photons CA. All these technical advancements allow us to realize the undetermined settings of MZI for S. Note that our work is device-dependent in the sense that we have to assume the generated state photons A and C is entangled and BSM between them is correctly performed. These assumptions have been independently verified with state fidelity measurements (up to 95\%) and two-photon interference experimental results shown in Appendix.

Details of the experimental setup are shown in Fig.~2. Femtosecond laser pulses (central wavelength 404~nm) are used to generate two photon pairs (central wavelength 808~nm) from two $\beta$-barium borate crystals (BBO1 and BBO2) via spontaneous parametric down-conversion (SPDC) processes\cite{zukowski1995entangling}. Specifically, from BBO1, photons C and A are generated in the $\ket{\psi^{+}}$ state in a non-linear configuration with walk-off compensations with half-wave plates (HWP) and BBO crystals with half thickness~\cite{PhysRevLett.75.4337}. From BBO2, photons S and T are generated in the $\ket{VH}_{ST}$ state in a collinear configuration. We use a polarization beam splitter (PBS) to separate photons S and T. The detection of photon T projects photon S onto a single-photon state. We couple all four photons into single-mode fibres for later manipulation and detection. Photon S passes through the polarization MZI which consists of one HWP orientated at an angle of $22.5^{\circ}$, a Soleil--Babinet compensator (SBC) and a quantum-CH gate~\cite{PhysRevLett.95.210504, PhysRevLett.95.210505, PhysRevLett.95.210506}. The SBC introduces a relative phase $\varphi$ between the $\ket{H}$ and $\ket{V}$ states. The quantum entanglement of photons C and A controls the setting of the CH gate. Before interfering with photon S on the first partial PBS (PPBS) of the CH gate, the polarization state of photon C is manipulated by a set of HWPs and quarter-wave plates (QWPs), QWP--HWP--QWP, to introduces the phase $\delta$. To obtain a successful operation of the CH gate, photons S and C have to arrive at PPBS1 simultaneously. We achieve that with several motorized translational stages mounted on single-mode fibre couplers. The BSM of photons C and A is implemented by overlapping them on PBS$_{\textrm{B}}$ and performing local polarization rotations on them with HWPs and QWPs. For details of the BSM and the cascade interferometers, see \textit{Appendix}.
    
In order to measure the correlation between photon S and photons C and A, Alice measures the individual polarization state of photon S along the angle $\theta_{1}$. The measurement of Bob is more complicated, as we need to perform joint measurements on photons C and A and to project them onto superpositions of Bell states. We realize this by rotating HWP${_2}$ with angle $\theta_{2}$ and overlapping photons C and A on PBS$_{B}$, as shown in Fig.~2. By doing so, Bob projects photons C and A onto a coherent superposition of the two orthogonal Bell states, $\cos\theta_{2} \ket{\phi^{-}}_{CA}+\sin\theta_{2}\ket{\psi^{+}}_{CA}$.

\section*{Correlation functions}  
Photon S can be a particle or a wave, or a quantum superposition of the two when we perform the corresponding Bell-state measurements on photons C and A, as shown in Eq. (2). Experimentally, we have four parameters to vary: the phase $\varphi$ of MZI, the phase $\delta$ between $\ket{w}$ and $\ket{p}$, the polarization-analyzer angle $\theta_{1}$ for photon S and the BSM-analyzer angle $\theta_{2}$ for photons C and A. We set $\delta$ at $\frac{\pi}{4}$ and then vary $\theta_{1}$, $\theta_{2}$ and $\varphi$ to measure the bipartite correlations functions $E_{S|CA}(\theta_{1},\theta_{2}, \varphi)$ between photon S, and photons C and A.
\begin{equation}
E(\theta_{1},\theta_{2},\varphi)=\frac{N^{++}-N^{+-}-N^{-+}+N^{--}}{N^{++}+N^{+-}+N^{-+}+N^{--}}
\end{equation}
where $N^{ij}$ is the coincidence count of Alice and Bob with measurement outcomes $i,j$ at the settings of $(\theta_{1},\theta_{2},\varphi)$.

\begin{figure*}
\includegraphics[width=17.8cm]{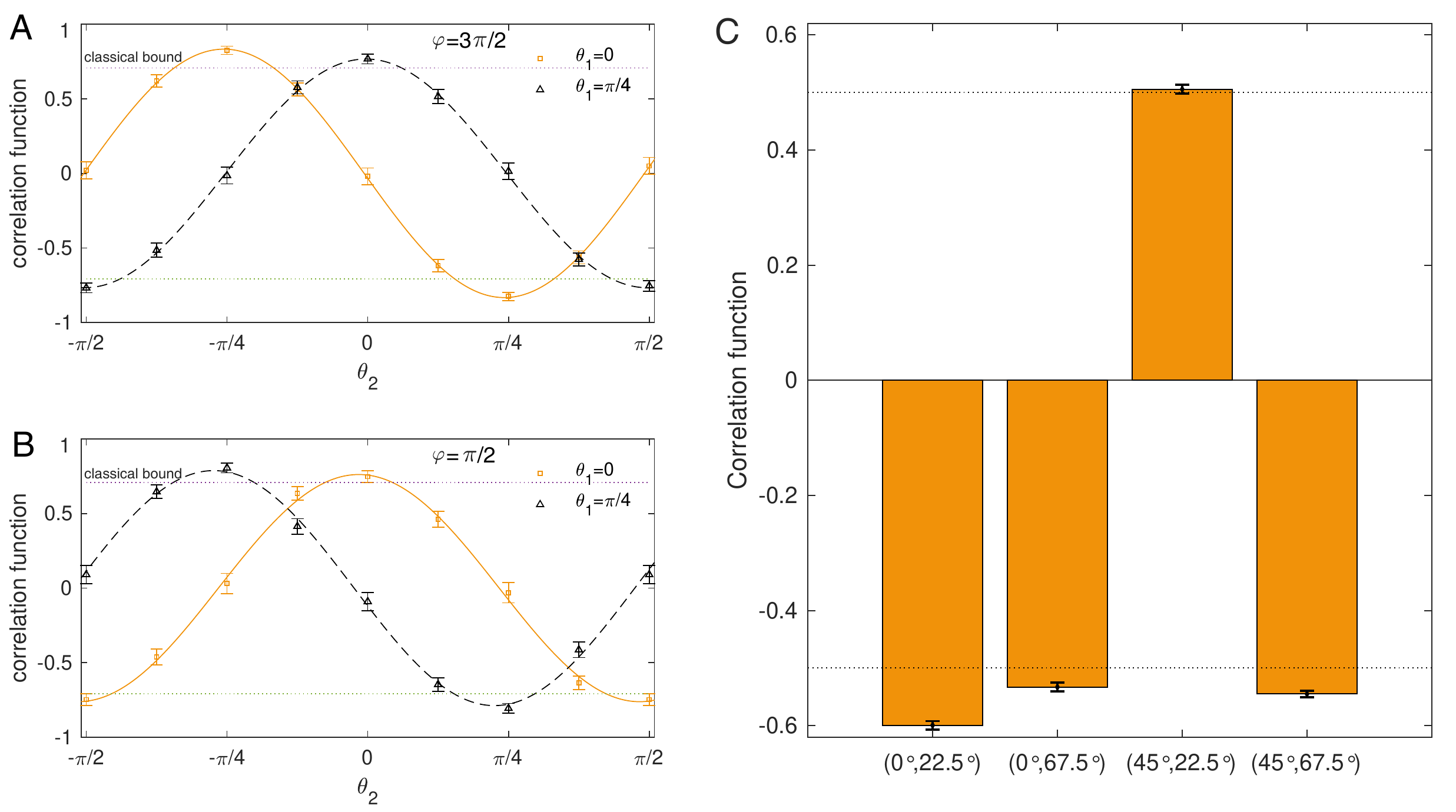}
\caption{ Experimental correlation functions $E_{S|CA}(\theta_{1},\theta_{2})$ of photon S, and photons C and A at MZI phase $\varphi=\frac{3\pi}{2}$ (\textit{A}) and $\varphi=\frac{\pi}{2}$ (\textit{B}). The experimental results of correlation functions for $\theta_{1}=0$ are shown in red circles and fitted with red solid curves, while those for $\theta_{1}=\frac{\pi}{4}$ are shown in black triangles and fitted with black dashed curves. (\textit{C}) Four correlation functions for violating CHSH inequality. $E(0,\frac{\pi}{8})=-0.5993\pm0.0072$, $E(0,\frac{3\pi}{8})=-0.5330\pm0.0076$, $E(\frac{\pi}{4},\frac{\pi}{8})=0.5056\pm0.0080$, $E(\frac{\pi}{4},\frac{3\pi}{8})=-0.5450\pm0.0053$. All the values are significantly above the classical bound of 0.5.}
\end{figure*}

In Fig.~3(\textit{A}) and (\textit{D}), we plot the ideal correlation function $E_{S|CA}$ at $\theta_{1}=0$ and $\pi/4$ as a function of $\theta_{2}$ and $\varphi$, respectively. According to the quantum state shown in Eq. (2), we have:
\begin{equation}
E(\theta_{2},\varphi)=\begin{cases}
\frac{1}{\sqrt{2}}[\cos{(2\theta_{2}+\frac{\pi}{4})}+\sin{\varphi}\cos{(2\theta_{2}-\frac{\pi}{4})}],\\
\theta_{1}=0;\\
\  \\
\frac{1}{\sqrt{2}}[\cos{(2\theta_{2}+\frac{\pi}{4})}-\sin{\varphi}\cos{(2\theta_{2}-\frac{\pi}{4})}],\\
\theta_{1}=\pi/4.\\
\end{cases}
\end{equation}
    
In Fig.~3(\textit{B}) and (\textit{C}), we show the theoretically expected and experimentally measured results for $\theta_{1}=0$, which agree well with each other. The parameters used to calculate the correlation shown in Fig.~3(\textit{B}) are based on the values obtained from independent experimental measurements. Note that the colour bars of Fig.~3(\textit{A}) is [-1, 1] and Fig.~3(\textit{B}) and (\textit{C}) are [-0.86, 0.86] respectively. This discrepancy is due to the visibility degradation caused by experimental imperfections, including high-order emission from the SPDC process and imperfect interference. In Fig.~3(\textit{D}), (\textit{E}) and (\textit{F}), we show the ideal, theoretically expected and experimentally measured results of the complementary correlation functions $E_{S|CA}$ at $\theta_{1}=\frac{\pi}{4}$ as a function of $\theta_{2}$ and $\varphi$.
    
Note that to obtain the experimental data shown in Fig.~3(\textit{C}) and (\textit{F}), we need to measure the coincidence counts of $N^{++}(\theta_{1},\theta_{2},\varphi)$, $N^{+-}(\theta_{1},\theta_{2},\varphi)$, $N^{-+}(\theta_{1},\theta_{2},\varphi)$ and $N^{--}(\theta_{1},\theta_{2},\varphi)$ at $\theta_{1}=0$ and $\frac{\pi}{4}$, respectively. For a standard Bell experiment, these four counts can be obtained in one run with two PBSs and four detectors owned by Alice and Bob, respectively. However, due to the design of our BSM, we can only obtain `+' outcomes in one run. Therefore, we measure the `-' outcome for $\theta_{2}$ by the `+' outcome for $(\theta_{2}+\pi/2)$. This is similar to the standard Bell experiment, in which Alice uses a wave-plate and PBS, and Bob uses a polarizer. Therefore, based on the definition of the correlation function in Eq. (3), we use the same coincidence counts to construct correlation functions for $E(\theta_{1},\theta_{2})$ and $E(\theta_{1},\theta_{2}+\pi/2)$. As a result, the data presented in Fig.~3(\textit{C}) and (\textit{F}) show symmetries. For instance, at $\varphi=\pi/2$, $E(0,-\pi/4)=-E(0,\pi/4)$. The measurement time of the data shown in Fig.~3(\textit{C}) and (\textit{F}) is one hour per data point, see \textit{Appendix} for details of the experiment results.

More importantly, when we set $\varphi=\frac{\pi}{2}$ and $\frac{3\pi}{2}$, we can violate a CHSH inequality with photon S and photons CA, as indicated by the vertical blue lines in Fig.~3(\textit{C}) and (\textit{F}). We show the correlation functions as a function of $\theta_{2}/2$ of HWP${_B}$ ($\theta_{2}$ from $-\frac{\pi}{2}$ to $\frac{\pi}{2}$) with $\varphi= \frac{3\pi}{2}$ and polarization projection angles of photon S set at $\theta_{1}=0$ (red squares) and $\theta_{1}=\frac{\pi}{4}$ (black triangles) in Fig.~4(\textit{A}), respectively. The red solid fitting curve ($\theta_{1}=0$) yields the visibility of $83.3\%$, whereas the black dotted curve ($\theta_{1}=\frac{\pi}{4}$) yields the visibility of $76.7\%$, both of which are above the classical bound of $\frac{1}{\sqrt{2}}$ required for violating Bell's inequality, as indicated by the dashed horizontal lines in Fig.~4(\textit{A}).

The maximum values of the $\theta_{1}=0$ correlation functions are $0.825\pm0.028$. The maximum values of the $\theta_{2}=\frac{\pi}{4}$ correlation functions are $0.768\pm0.033$. We then set the phase of the MZI to $\varphi= \frac{\pi}{2}$ and measured the complementary correlations. The results are shown in Fig.~4(\textit{B}), from which we obtain the solid $\theta_{1}=0$ fitting curve, yielding the visibility of $76.2\%$, whereas the dotted $\theta_{2}=\frac{\pi}{4}$ curve yields the visibility of $78.8\%$. The maximum values of the $\theta_{1}=0$ correlation functions (red solid curve) are $0.747\pm0.039$. The maximum values of the $\theta_{2}=\frac{\pi}{4}$ correlation functions (black dotted curve) are $0.806\pm0.032$.
    
To obtain a significant violation of the CHSH inequality, we fixed the power of the 404-nm laser at 0.28~W and measured eight hours per data point. From these results (shown in Fig.~4(\textit{C})), we obtain the following four correlations: $E(0,\frac{\pi}{8})=-0.5993\pm0.0072$, $E(0,\frac{3\pi}{8})=-0.5330\pm0.0076$, $E(\frac{\pi}{4},\frac{\pi}{8})=0.5056\pm0.0080$, $E(\frac{\pi}{4},\frac{3\pi}{8})=-0.5450\pm0.0053$. This gives the Bell parameter  $S=|E(\theta_{1},\theta_{2})+E(\theta_{1},\theta'_{2})-E(\theta'_{1},\theta_{2})+E(\theta'_{1},\theta'_{2})|=2.1829\pm0.0302$, which violates the CHSH inequality by more than six standard deviations.

\section*{Conclusion}
In conclusion, we realize an experimental scenario, in which the setting of the interferometer for photon S is controlled by the quantum state of photon C. Since C and A are entangled, the individual state of C is undefined. Consequently, the setting of the MZI for S and the wave-particle property of S is not well defined at any instance of the experiment, not even after the measurements of photons S, C and A. By using the controlled-Hadamard gate, we establish the entanglement between the wave/particle state of photon S and the Bell states of photon C and A, as shown in Eq. (2). The control of the wave-particle properties of photon S via photons CA is manifested via the three--photon correlation functions shown in Fig.3. The quality of such control is captured by the high-confidence violation of Bell inequality. Our work can be viewed as entangled entanglement control of wave-particle duality.

If an HV description is valid, then all individual results can be predicted. As the settings are set before each run, the $\lambda-$independence is not enforced, making the arguments of Ref.~\cite{PhysRevLett.114.060405} inapplicable. However, this still does not provide the wave--particle objectivity, because the observed phenomena do not allow to introduce the notion of `wave' and `particle' to describe photon S under the device dependent assumptions, which have been experimentally verified.

Our work reveals the very essence of complementarity. According to Bohr, quantum phenomena observed on a system depend on the entire experimental contexts "which serve to define the conditions under which the phenomena occur"~\cite{bohr1996discussion}. Our work shows that the definiteness of the context is not only sufficient but also necessary condition to define quantum phenomena: the wave-particle objectivity of the individual system is ill defined when the interferometric setting for the system is not defined.
	
\section{acknowledgments}
The authors thank A. Zeilinger for helpful discussions. This research is supported by the National Key Research and Development Program of China (2017YFA0303704, 2019YFA0308700), National Natural Science Foundation of China (Grants No. 11690032), NSFC-BRICS (No. 61961146001), Leading-edge technology Program of Jiangsu Natural Science Foundation  (BK20192001), and the Fundamental Research Funds for the Central Universities.

	%
	%
\appendix
\setcounter{figure}{0}
\renewcommand{\figurename}{Fig.}
\renewcommand{\thefigure}{A\arabic{figure}}
\renewcommand{\tablename}{Table.}
\renewcommand{\thetable}{A\arabic{table}}
\setcounter{equation}{0}
\renewcommand{\theequation}{A\arabic{equation}}
\renewcommand{\thesubsection}{A\arabic{subsection}}

\section*{Appendix}
\subsection{I. Hidden Variable Description of Objectivity}
Hidden-variable (HV) models strive  to explain or even to remove the features of quantum theory that run contrary to the classical intuition while reproducing
its experimental predictions. A HV model contains two elements. Given the classical settings $(A,B,\ldots)$ of the measurement devices it produces  a conditional probability distribution of
the observable quantities $(a,b,\ldots)$ that depends on the value  of HV $\Lambda$, $p(a,b,\ldots|\Lambda)$. The HV itself, who may have an arbitrary unspecified structure is unknown, and the HV model contains some probability
distribution $p(\Lambda)$.  Predictions for the observed probabilities $p(a,b,\ldots)$ are obtained by an appropriate
integration or summation over $\Lambda$.

First we briefly summarize the standard properties of HV models. The details can be found in, e.g., Ref.~\cite{brandenburger2008a}.
Viable HV models must be adequate \cite{brandenburger2008a}, i.e. reproduce the empirical data that  agrees with the predictions of quantum theory, $q(a,b,\ldots)=p(a,b,\ldots)$.
Imposing classical requirements (determinism, versions of
locality, etc.) on HV models constrains the resulting
probability distributions. The requirements may be incompatible, failing to produce a single $p(a,b,\ldots)$. Famous paradoxes of quantum mechanics are expressed
as failures the HV probability distributions cannot match the quantum $q(a,b,\ldots)$.

Determinism can be formally represented in several forms.  Either strong or weak determinism express the idea
that once the relevant HV are set, the outcomes of all specified measurements are pre-determined.

Parameter independence is a statement that conditional on the value of the hidden variable, the outcome of any  measurement probabilistically depends only on the settings of the measurement itself and not on the other measurements. In terms of conditional probabilities, it reads:
\be
p(a|A,B,C,\ldots, \Lambda)=p(a|A,\lambda).
\ee

$\lambda$-Independence is a property of HV model that the process determining the value of the hidden variable is independent of the measurement settings. It reads:
\be 
p(\Lambda|A,B,C,\ldots)=p(\Lambda|A',B',C',\ldots).
\ee
Other properties are derived from these primitives. 

We operationally define particle-like and wave-like properties according to the behavior in an open (respectively closed),
balanced MZI.  A particle in an open  interferometer is insensitive to the phase shift in one of the arms,
while a wave in a closed MZI  shows interference.    Hence in the basis that we conventionally label by the outcomes $s=0,1$, the `particle' and the `wave' statistics   are given by
\be
e_p=\left(\half,\half\right), \qquad e_w=\left(\cos^2\tfrac{\varphi}{2},\sin^2\tfrac{\varphi}{2}\right), \ee
respectively.

The original delayed choice experiment is described by the top two wires in Fig. 1(\textit{A}) of the main text. The MZI being open or closed is determined by the controlled-Hadamard gate where the control can be either $(c=0)$ (open MZI)
or $(c=1)$ (closed MZI). A quantum-controlled  Hadamard gate allows for the interferometer to be in an arbitrary superposition of being open and closed. Any HV theory that imposes the notion of wave-particle objectivity
has to use the probability distributions
\be
p(s|c=0,\Lambda)=e_p, \qquad p(s|c=1,\Lambda)=e_w, \qquad  \forall \Lambda.  \label{defHV}
\ee

However, the current scheme is different. Having the control qubit entangled with the qubit $A$, the measurements  by Bob in the basis
$\ket{\pm;\theta_2}_{CA}= \cos\theta_2\ket{\phi^{-}}_{CA}\pm i\sin\theta_2\ket{\psi^{+}}_{CA}$  generate the wave and particle statistics,
\be
q(s|+;\theta_2)=e_w, \qquad q(s|-;\theta_2)=e_p.
\ee
If a HV description is valid, then all individual results can be predicted beforehand.  However, apart from the special values  $\theta_2=\tfrac{\pi}{4}$,
there is no definite state of the MZI being open or closed. The state of C is mixed, with either outcome being equiprobable.
The wave- and particle- like statistics are exhibited in either case, conditioned on the outcome of the partial Bell measurement.
In this situation objective properties ``being a wave'' or ``being a particle''  (as summarized by Eq.~\eqref{defHV})  cannot be introduced.

\subsection{II. Cascade Hong-Ou-Mandel Interference}

In our experiment, we use two-photon interference to realize the control-Hadmard (CH) gate and the Bell state measurement (BSM), as shown in Fig. 1(\textit{C}) of the main text.

\textit{Interference in CH gate.} CH gate consists of two W gates and one Control-Z (CZ) gate.  In the CZ gate, photon C interferes with  photon S on a partial polarization beam splitter (PPBS1) with perfect transmission $T_{h}$ for horizontal polarization and transmission $T_{v} = 1/3$ for vertical polarization in both output modes \cite{PhysRevLett.95.210504,PhysRevLett.95.210505,PhysRevLett.95.210506}. Single-mode fiber coupler of photon C is mounted on a motorized translation stages 1 to introduce time delay when we scan the HOM interference pattern of photons S,C. Photons A and T are detected directly after being collected into the coupler, while photons S, C are sent into the Mach-Zehnder interferometer (MZI). We count the four-fold coincidence counts of photons S, C, A, T in the polarization state $\ket{VVHH}_{SCAT}$ as stage 1 moves. Fig. A1 shows the coincidence counts when $\varphi=0$ which gives a contrast 0.848. Then the location of stage 1 is fixed. Note that when we set different phases $\varphi$ in MZI, the thickness of the Soleil-Babinet compensator changes, modifing the location of HOM dip.

\begin{figure}
    \centering
    \includegraphics[width=8cm]{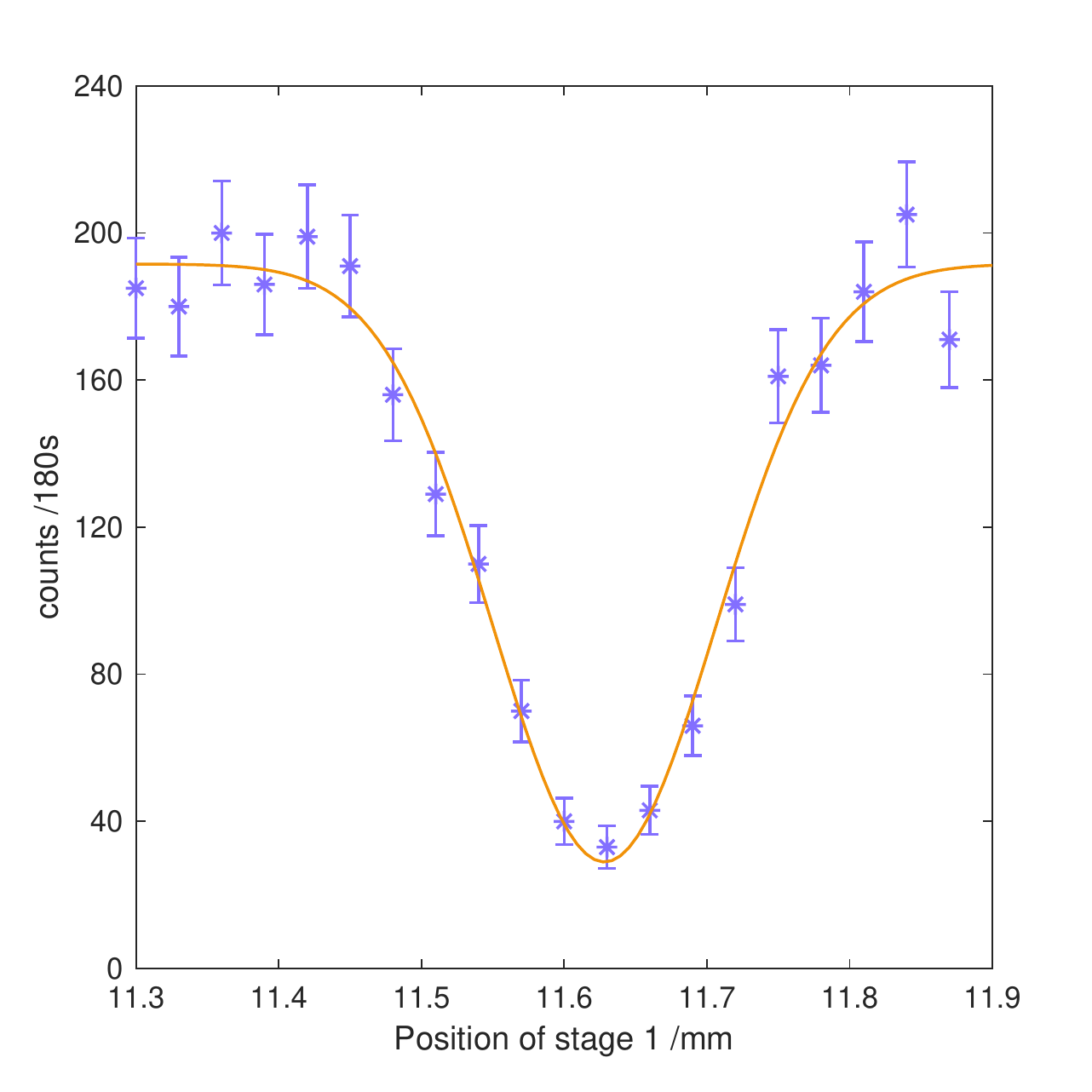}
    \caption{\label{fig:wide} HOM interference result when phase $\varphi=0$. Y axis represents the four-fold coincidence counts of photons S,C,A,T in the polarization state $\ket{VVHH}_{SCAT}$. The integration time of each point is 180~s. The fitting curve gives the minimum value at 11.63~mm where the contrast is 0.848.}
\end{figure}

\textit{Interference in BSM.} BSM is realized with one a half-wave plate ($\textrm{HWP}_{2}$) used for choosing the measurement angle $\theta_{2}$ and a polarizing beam splitter ($\textrm{PBS}_{B}$). $\theta_{2}$ is two times the setting angle of $\textrm{HWP}_{2}$. To find the HOM dip on $\textrm{PBS}_{B}$, we mount the single-mode fiber coupler of photons A on another motorized translation stage 2 and scan it. We send photon C from CH gate and photon A into BSM module and accumulate the two-fold coincidence counts of photons C,A. Photon S is blocked in this process. When photons C,A arrive on PBS at the same time, we obtain the Bell state, $\ket{\psi}_{CA}=\frac{\ket{DA}+\ket{AD}}{\sqrt{2}}$, by detect them in $\ket{D}$/$\ket{A}$ basis. To show the coherence between $\ket{DA}$ and $\ket{AD}$, we measure the coincidence counts in the four settings: $\ket{DA}$, $\ket{DD}$, $\ket{AA}$ and $\ket{AD}$. The results are shown in Fig. A2. It is clear to see that, at the position of about 14.42 mm, $\ket{DA}$ and $\ket{AD}$ coincidence counts show maximum while $\ket{DD}$ and $\ket{AA}$ coincidence counts are minimum, confirming the entanglement of photon C,A in state $\ket{\psi}_{CA}$. For above measurements, we set $\theta_{2}$ and phase $\delta $ to be 0.

\subsection{III. Correlations and the violation of Bell's inequality}
In this section, we firstly show the raw four-fold coincidence counts ($C_{i}$) between photons S,C,A,T in Fig. A3. Photon S is projected into both $\ket{D}$/$\ket{A}$ and $\ket{D}$/$\ket{A}$ bases, while photons C and A are projected into $\ket{D}$/$\ket{A}$ basis. Photon T is the trigger and is projected onto $\ket{H}$. The subscript $i$ in $C_{i}$ is the polarization projection state for photon S.

From these coincidence counts, we obtain the correlation functions shown in Fig. 3(\textit{C}) and (\textit{F}) of the main text. Here we show the specific values of the correlation functions together with fitting curves in Fig. A4. Note that the data presented in Fig. A4\textbf{a} and \textbf{b} are identical to those shown in the Fig. 3(\textit{C}) and (\textit{F}) of the main text.

After we obtain these good correlations, we proceed to the violation of Bell's inequality. In order to obtain the significant violation, we reduce the pump power to reduce the high order emissions from SPDC sources and integrate eight hours for each data point. Table S1 shows coincidence counts used in correlation function calculation. With the data from Table. A1, we obtain the four correlation parameters: $E(0,\frac{\pi}{8})=-0.5993\pm0.0072, E(0,\frac{3\pi}{8})=-0.5330\pm0.0076, E(\frac{\pi}{4},\frac{\pi}{8})=0.5056\pm0.0080, E(\frac{\pi}{4},\frac{3\pi}{8})=-0.5450\pm0.0053$, as shown in Fig. 4(\textit{C}) of the main text.

\begin{table}
\centering
\caption{Coincidence Counts}
\begin{tabular}{|c|c|c|c|c|c|c|c|}
    \hline
    \multicolumn{3}{|c|}{}& \multicolumn{4}{c|}{$\theta_{2}$}\\
    \cline{4-7}
    \multicolumn{3}{|c|}{} & $22.5^{\circ}$ & $-67.5^{\circ}$ & $67.5^{\circ}$ & $-22.5^{\circ}$ \\ \hline
    \ & $0^{\circ}$ & H & 577 & 2164 & 454 & 2521 \\ \cline{3-7}
    $\theta_{1}$ & \ & V & 2302 & 542 & 1884 & 888 \\ \cline{2-7}
    \ & $45^{\circ}$ & D & 2162 & 658 & 402 & 2524 \\ \cline{3-7}
    \ & \ & A & 692 & 1949 & 1911 & 904 \\ \hline
\end{tabular}\\
\end{table}

\subsection{IV. Summary of recent experiments on wave-particle properties of photons}
\quad A summary of recent experiments on wave-particle properties of photons is shown in Table. A2.\\

\begin{enumerate}
\item In ref 23 [Peruzzo, et al. Science 338, 634], the control photon is prepared in a well-defined state before the quantum control operation. Therefore, the wave, particle, and their superposition properties are defined by the well-defined single-qubit measurement results of the control photon. \\
\item In ref 24 [Kaiser, et al. Science 338, 637], the quantum state of control photon is well-defined after the single-photon polarization measurement. Therefore, the wave, particle, and their superposition properties are defined by the well-defined single-qubit measurement results of the control photon.\\ 
\item In ref 30 [Wang, et al. Nat. Photon. 13, 872], the quantum state of control photon is well-define after the two-photon product-state measurement. Therefore, the wave, particle, and their superposition properties are defined by the well-defined two-qubit product measurement results. \\
\item In the present work, the arrangement of the interferometer is not defined, not even in principle. This is because the control photon is one part of a maximally entangled bipartite state (formed by photons C and A). Therefore, the individual quantum state of photon C is undefined, which implements the undefined arrangement of the MZI for photon S. This scenario is feasible thanks to the unique quantum control with entangled entanglement between three photons.\\
\end{enumerate}

\begin{figure*}
    \centering
    \includegraphics[width=14cm]{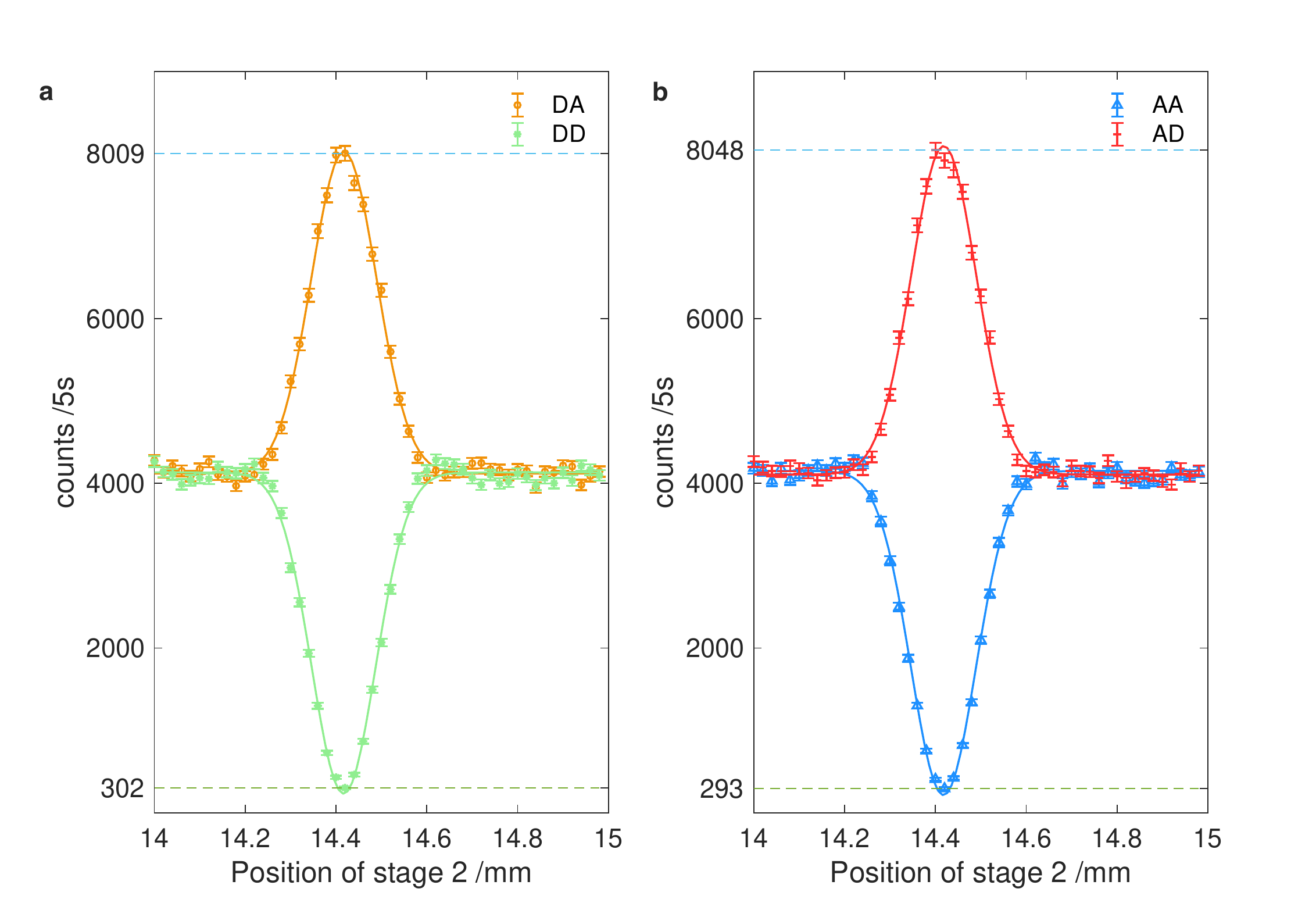}
    \caption{\label{fig:wide} 2-fold coincidence counts of photon C, A. The integral time of each point is 5~s. \textbf{a} Coincidence counts on basis $\ket{DA}$ and $\ket{DD}$. \textbf{b} Coincidence counts on basis $\ket{AA}$ and $\ket{AD}$. The counts show positive correlation in basis $\ket{DA}$ and $\ket{AD}$ confirming the entanglement between photon C, A.}
\end{figure*}

\begin{figure*}
    \centering
    \includegraphics[width=18cm]{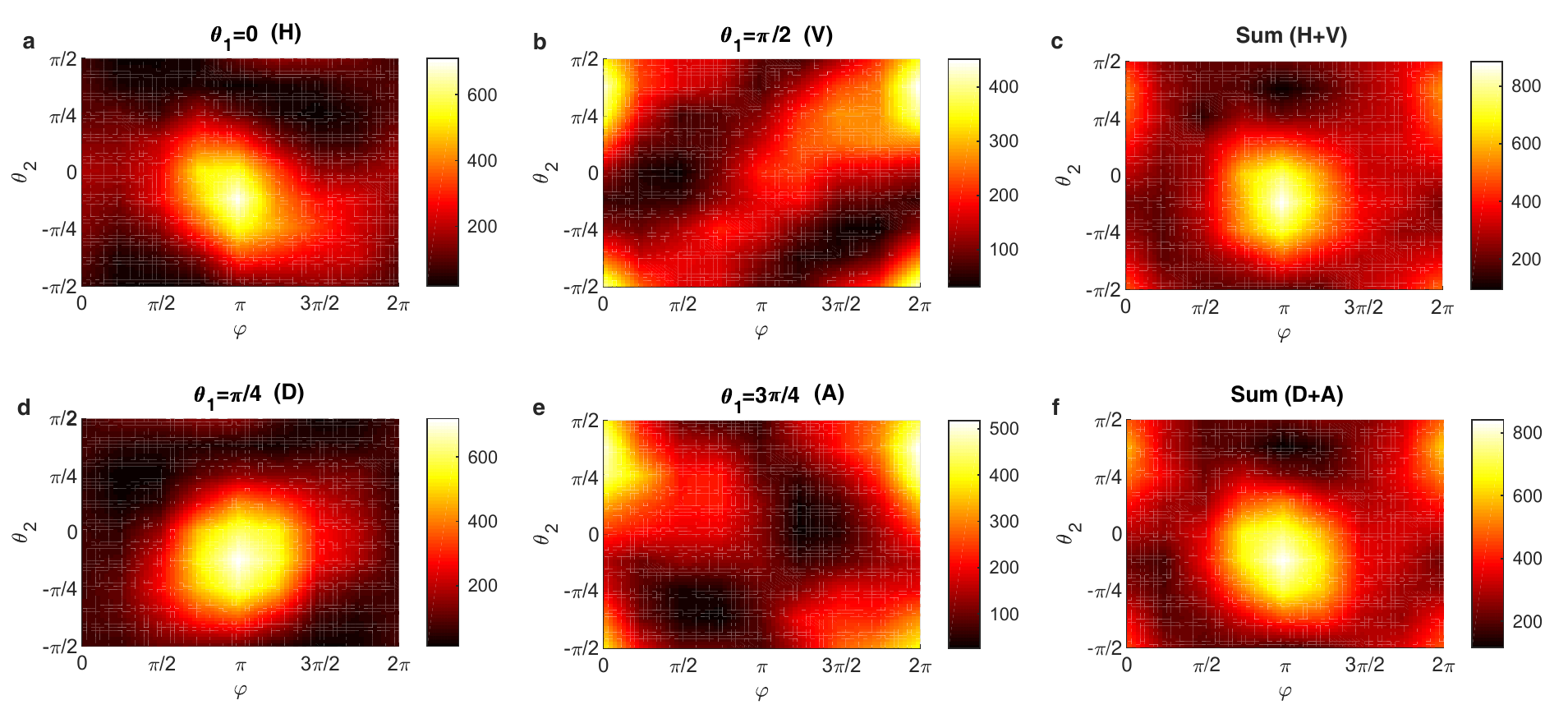}
    \caption{\label{fig:wide} Experimental result of raw four-fold coincidence counts $C_{i}$ of photons S,C,A,T. (a) $C_{H}$ (b) $C_{V}$ (c) $C_{H}+C_{V}$ (d) $C_{D}$ (e) $C_{A}$ (f) $C_{D}+C_{A}$. The subscript $S$ in $C_{i}$ represents the polarization projection state for photon S.}
\end{figure*}

\begin{figure*}
    \centering
    \includegraphics[width=18cm]{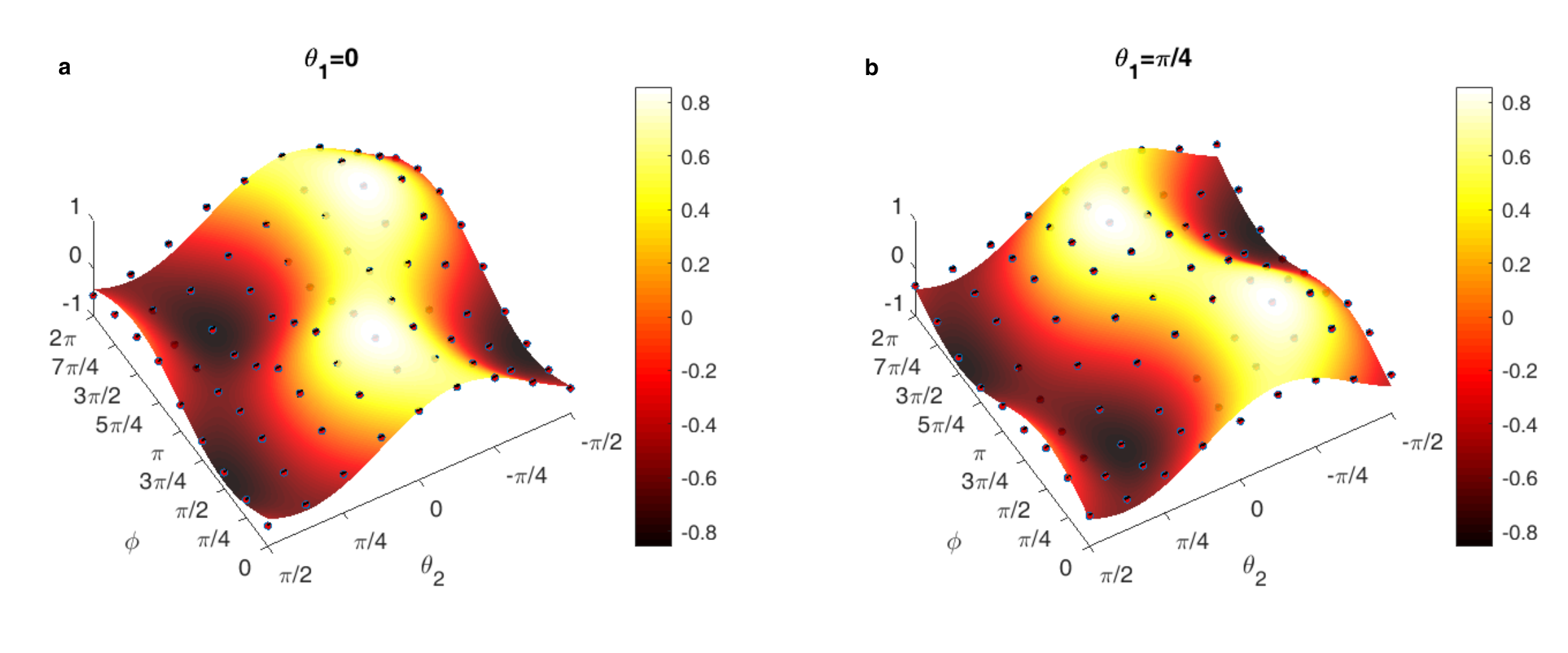}
    \caption{ Correlation parameter $E(\theta_{1},\theta_{2},\varphi)$. Alice changes the phase $\varphi$ in the MZI to obtain different final states $\ket{\Psi^{f}}$. When $\phi=\pi/2, 3\pi/2$, $\ket{\Psi^{f}}$ is the maximum entangled state and the correlation function (changing $\theta_{2}$) gives the largest visibility. Bob changes the Bell state measurement (BSM) angle $\theta_{2}$ from $-\pi/2$ to $\pi/2$ while Alice measures the polarization of photon S on $\ket{H},\ket{V}$ basis ($\theta_{1}=0^{\circ}$), shown in \Bo{a},  and $\ket{D},\ket{A}$ basis ($\theta_{1}=45^{\circ}$), shown in \Bo{b}. }
\end{figure*}

\begin{table*}
\makeatletter\def\@captype{table}\makeatother
\centering
\caption{Summary of recent experiments on wave-particle properties of photons}
\begin{tabular}[c]{|c|c|c|c|c|}
    \hline
    Refs & \makecell{Methods used to \\control the MZI} & \makecell{ \hspace{0.5em}Photon\hspace{0.5em} \\ \hspace{0.5em}number \hspace{0.5em}} & \makecell{Interferometer\\ number} & Implications\\
    \hline
    \makecell{Jacques, V. et al.\\ Science 315,\\ 966-968 (2007).} & Classical bits & 1 & 1 & \makecell[l]{\hspace{0.5em}The wave or particle prop-\\\hspace{0.5em}erties are defined by exper-\\\hspace{0.5em}imental settings controlled\\ \hspace{0.5em}by classical bits under Ein-\\\hspace{0.5em}stein's locality condition.}\\
    \hline
    \makecell{Jacques, V. et al.\\ Phys. Rev. Lett. 100,\\ 220402 (2008).} & Classical bits & 1 & 1 & \makecell[l]{\hspace{0.5em}The wave or particle prop-\\\hspace{0.5em}erties are defined by exper-\\\hspace{0.5em}imental settings controlled \\\hspace{0.5em}by classical bits under Ein-\\\hspace{0.5em}stein's locality condition.}\\
    \hline
    \makecell{Peruzzo, A., et al.\\ Science 338,\\ 634-637 (2012).} & \makecell{Well-defined\\ single-qubit states} & 2 & \makecell{7 (3 interferometer \\for control; 2 for\\ state preparation; 2\\ for measurement)} & \makecell[l]{\hspace{0.5em}The wave, particle, and \\\hspace{0.5em}their superposition proper-\\\hspace{0.5em}ties are defined by exper-\\\hspace{0.5em}imental settings controlled\\ \hspace{0.5em}by well-defined single-qubit\\ \hspace{0.5em}measurement results.}\\
    \hline
    \makecell{Kaiser, F. et al.\\ Science 338,\\ 637-640 (2012).} & \makecell{Well-defined\\ single-qubit states} & 2 & 1 & \makecell[l]{\hspace{0.5em}The wave, particle, and \\\hspace{0.5em}their classical mixture \\ \hspace{0.5em}properties are defined \\\hspace{0.5em}by experimental settings\\ \hspace{0.5em}controlled by well-defined\\ \hspace{0.5em}single-qubit measurement\\ \hspace{0.5em}results under Einstein's\\ \hspace{0.5em}locality condition.}\\
    \hline
    \makecell{Wang, K. et al.,\\ Nat Photon 121,\\ 1-6 (2019).} & \makecell{Two-qubit product states} & 3 & 2 & \makecell[l]{\hspace{0.5em}The wave, particle, and\\ \hspace{0.5em}their quantum superposi-\\\hspace{0.5em}tion properties are defined\\ \hspace{0.5em}by experimental settings\\ \hspace{0.5em}controlled by local two-\\ \hspace{0.5em}qubit measurement results\\\hspace{0.5em}under Einstein's locality\\\hspace{0.5em}condition.}\\
    \hline
    Present work & \makecell{Entangled two-qubit states} & 3 & 3 & \makecell[l]{\hspace{0.5em}We demonstrate that unde-\\\hspace{0.5em}fined setups reveal undefined\\ \hspace{0.5em}properties of single photons. \\ \hspace{0.5em}Therefore, the very formula-\\\hspace{0.5em}tion of the wave-particle\\ \hspace{0.5em}objectivity becomes interna-\\\hspace{0.5em}lly inconsistent.}\\
    \hline
\end{tabular}
\end{table*}

\clearpage
%

\end{document}